%% file: eprint.tex
\documentclass[12pt]{article}
\usepackage{graphicx}
\usepackage{subfigure, wrapfig}

\newcommand\pubnumber{SNSN-323-63}
\newcommand\pubdate{\today}

\def\lbl{Institute for Nuclear and Particle Astrophysics, Lawrence Berkeley Laboratory, Berkeley, CA 94720, USA}
\def\kit{KIT Center Elementary Particle and Astroparticle Physics, Karlsruhe Institute of Technology, 76021 Karlsruhe, Germany}

\def\Title#1{\begin{center} {\Large #1 } \end{center}}
\def\Author#1{\begin{center}{ \sc #1} \end{center}}
\def\Address#1{\begin{center}{ \it #1} \end{center}}

\newcommand\pubblock{\rightline{\begin{tabular}{l} \pubnumber\\
         \pubdate  \end{tabular}}}
\newenvironment{Abstract}{\begin{quotation}  }{\end{quotation}}
\newenvironment{Presented}{\begin{quotation} \begin{center} 
             PRESENTED AT\end{center}\bigskip 
      \begin{center}\begin{large}}{\end{large}\end{center} \end{quotation}}

\input econfmacros.tex

\begin{document}
\begin{titlepage}
\pubblock

\vfill
\Title{Direct Neutrino Mass Experiments}
\vfill
\Author{Susanne Mertens}
\Address{\lbl}
\Address{\kit}
\vfill
\begin{Abstract}
With a mass at least six orders of magnitudes smaller than the mass of an electron -- but non-zero -- neutrinos are a clear misfit in the Standard Model of Particle Physics. On the one hand, its tiny mass makes the neutrino one of the most interesting particles, one that might hold the key to physics beyond the Standard Model. On the other hand this minute mass leads to great challenges in its experimental determination. Three approaches are currently pursued: An indirect neutrino mass determination via cosmological observables, the search for neutrinoless double $\beta$-decay, and a direct measurement based on the kinematics of single $\beta$-decay. In this paper the latter will be discussed in detail and the status and scientific reach of the current and near-future experiments will be presented.
\end{Abstract}
\vfill
\begin{Presented}
NuPhys2015, Prospects in Neutrino Physics \\
Barbican Centre, London, UK,  December 16--18, 2015
\end{Presented}
\vfill
\end{titlepage}
\def\thefootnote{\fnsymbol{footnote}}
\setcounter{footnote}{0}

\section{Introduction}

The discovery of neutrino oscillations has proven that neutrinos have a non-zero mass~\cite{pdg}. Yet, the absolute neutrino mass scale is still unknown since oscillation experiments are only sensitive to the squared mass differences of the three neutrino mass eigenstates $m_{\nu_i}$. The knowledge of the neutrino mass is crucial both for Particle Physics and for Cosmology. It will be an essential ingredient to answering the question of the neutrino mass generation mechanism, and an important input parameter to reduce degeneracies in cosmological models.

Until more stringent results from laboratory-based experiments are established, cosmological observations themselves provide powerful probes of the neutrinos mass. Current limits based on a combination of cosmological probes set limits of $m_{\nu} = \sum_i m_{\nu_i}<120$~meV (95\% C.L.)~\cite{cosmo}. Future experiments aim to reach a precision of $\sigma({m_{\nu} }) = 17$~meV~\cite{cosmofuture}. It is important to note, however, that these results will depend on the underlying cosmological model.

Another sensitive probe of the neutrino mass is the search for neutrinoless-double $\beta$-decay ($0\nu\beta\beta$). Here, one exploits the fact, that the half-life of the decay depends on the so-called Majorana neutrino mass $m_{\beta\beta}  = |\sum_i U^2_{ei}m_{\nu_i}|$.  Current best limits are at $m_{\beta\beta} = 120 - 250$~meV~\cite{0nbb} and, thanks to their scalability, future $0\nu\beta\beta$ experiments plan to reach sensitivities down to $m_{\beta\beta} \approx 25$~meV~\cite{Avi08}. 

The least model-dependent technique is solely based on the kinematics of single-$\beta$-decay. Here, the impact of the so-called effective electron (anti-)neutrino mass $m^2_{\nu_e}  = \sum_i |U_{ei}|^2 m_{\nu_i}^2$ is a reduction of the endpoint energy and a distortion of the spectrum close to the endpoint. Near-future experiments are designed to reach a sensitivity of $m_{\nu_e} = 200$~meV (90\% C.L.)~\cite{Drex13}, probing the entire regime in which the neutrino mass eigenstates are quasi-degenerate. New ideas are being explored to push the sensitivity beyond this value to the inverted or normal hierarchical neutrino mass regime.

\section{Kinematic determination of the neutrino mass}
\begin{figure}[]
\begin{center}
\includegraphics[width = 0.85\textwidth]{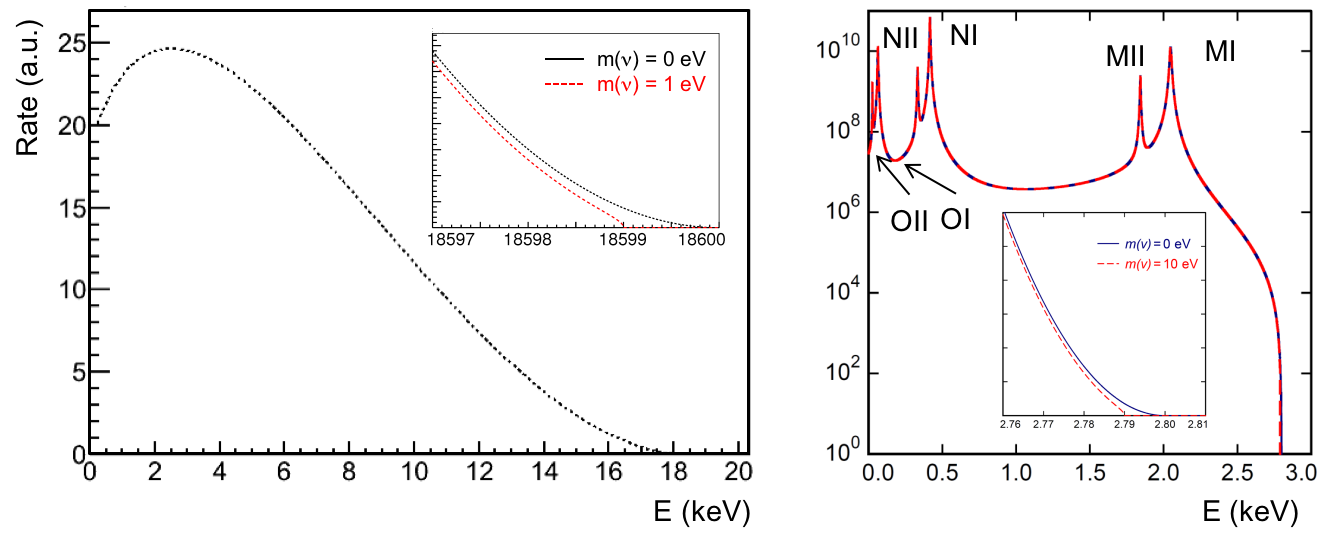}
\caption{a: Tritium $\beta$-decay spectrum. b: Holmium-163 electron-capture spectrum. The insets depict a zoom into the endpoint region and demonstrate the impact of a finite effective electron neutrino mass.}
\label{fig:spectra}
\end{center}
\end{figure}

For a kinematic determination of the neutrino mass generally a single $\beta$-decay is considered. Neglecting the small recoil of the heavy daughter nucleus, only the emitted electron and neutrino statistically share the energy released in the decay. The electron, however, can never obtain the entire decay energy, since the neutrino takes away at least the amount of energy that corresponds to its mass. Consequently, the maximum electron energy is reduced and the spectrum is distorted in the close vicinity of the spectrum's endpoint $E_0$, see figure~\ref{fig:spectra}. 

\subsection{Isotopes under consideration}
The classical isotope in the field of neutrino mass measurement is tritium (${}^{3}$H).  ${}^{3}$H has an endpoint of 18.6~keV and decays with a half-live of 12.3~years via a super-allowed $\beta$-decay to helium-3 (${}^{3}$He). Short half-life and low endpoint are preferable, since in this case the total decay rate per amount of isotope and the relative fraction of events in the region of interest are maximized.

Holmium-163 (${}^{163}$Ho) constitutes a new player  in the field. It has an endpoint of about 2.8~keV~\cite{eli15} and decays with a half-life of 4570~years via electron-capture to dysprosium-163 (${}^{163}$Dy). In this case there is no electron in the final state and instead of an anti-neutrino a neutrino is emitted. Here, the decay energy is shared between the neutrino and the excitation of the daughter nucleus ${}^{163}$Dy, which in turn decays via the emission of X-rays and Auger and Coster-Kronig electrons.

\section{Current experimental efforts}
Independent of the isotope, a major experimental requirement is an excellent energy resolution  of about 2~eV @ $E_0$ in order to resolve the spectral distortion that only extends over an energy range of few eV at the endpoint. To allow for a measurement as close as possible to the endpoint where the signal rate is small, but the neutrino mass signal is large, an extremely low background level is mandatory.

\subsection{The KATRIN Experiment}
The Karlsruhe Tritium Neutrino (KATRIN) experiment is a large-scale tritium-$\beta$-decay experiment~\cite{KAT04}. It is currently being commissioned at the Karlsruhe Institute of Technology, Germany. KATRIN is designed to achieve a neutrino mass sensitivity of 200~meV (90\% C.L.) after 3 full-beam years of measurement time. 

\subsubsection{Working principle}
\begin{figure}[]
\begin{center}
\includegraphics[width = 0.8\textwidth]{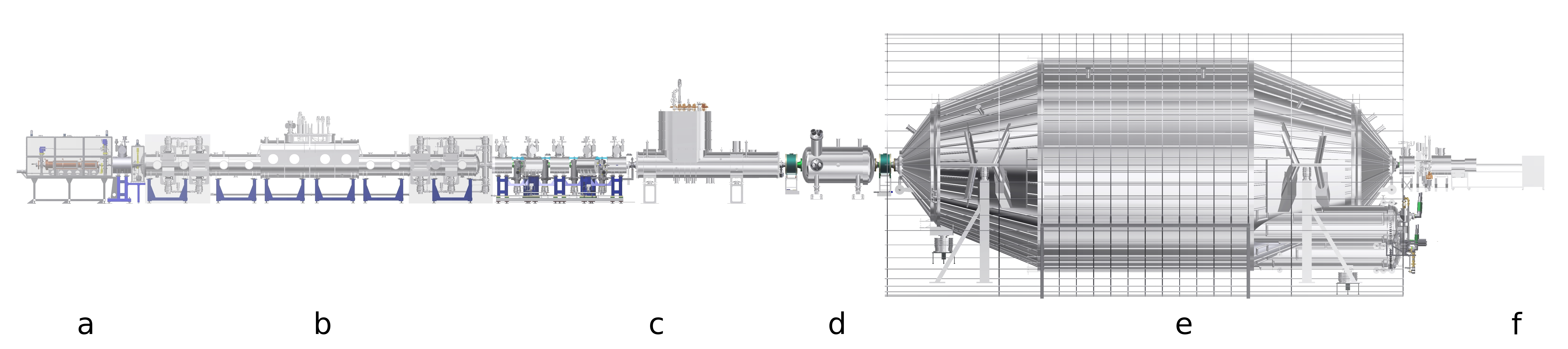}
\caption{Main components of the 70-m-long KATRIN experimental setup. a: rear section, b: windowless gaseous tritium source, c: differential and cryogenic pumping section, d: pre-spectrometer, e: main spectrometer, f: focal plane detector.}
\label{fig:KATRIN}
\end{center}
\end{figure}

Tritium of very high isotopic purity ($>95\%$) is injected through capillaries into the windowless gaseous tritium source (WGTS) tube, see figure~\ref{fig:KATRIN}. The $\mathrm{T}_2$ molecules then diffuse over a distance of 5~m to both ends of the WGTS. With about 30~$\mu$g of tritium present in the WGTS at all times, an ultra-high and stable decay rate of $10^{11}$ decays/s is achieved.

The WGTS beam tube is situated in a magnetic field, which is oriented in beam direction. All $\beta$-electrons that are emitted in the forward direction are guided along the field lines towards the spectrometers. On the way from the WGTS to the spectrometers the flow of tritium has to be reduced by 14 orders of magnitude to avoid tritium-related background in the spectrometer section. This large suppression factor is achieved by a combination of differential and cyrogenic pumping.

The spectrometers work as a electrostatic filters allowing only those electrons with enough kinetic energy to be transmitted; electrons with less kinetic energy than the filter potential will be electrostatically reflected and are absorbed at the rear end. The high-energy transmitted electrons reach a focal-plane detector where they are counted. By varying the filter potential and counting the transmitted electrons for each setting, the integral tritium spectrum is determined.

In addition to electrostatically filtering the electrons, the spectrometer also aligns the electron momenta via the magnetic gradient force. This combination of magnetic adiabatic collimation combined with electrostatic filtering is called MAC-E Filter principle~\cite{Lob85, Pic92} and allows for high energy resolution with large angular acceptance. For the electromagnetic design of the KATRIN the maximal acceptance angle is $51^{\circ}$ and the sharpness of the electrostatic filter (or energy resolution) is 0.93~eV.

\subsubsection{Status and Sensitivity}
Since September 2015 all KATRIN components are on-site at KIT. The windowless gaseous tritium source, cryogenic and differential pumping section are currently being commissioned and integrated~\cite{Luk11,Bab12}. 

\begin{figure}
  \centering
  \subfigure[]{\includegraphics[width = 0.41\textwidth]{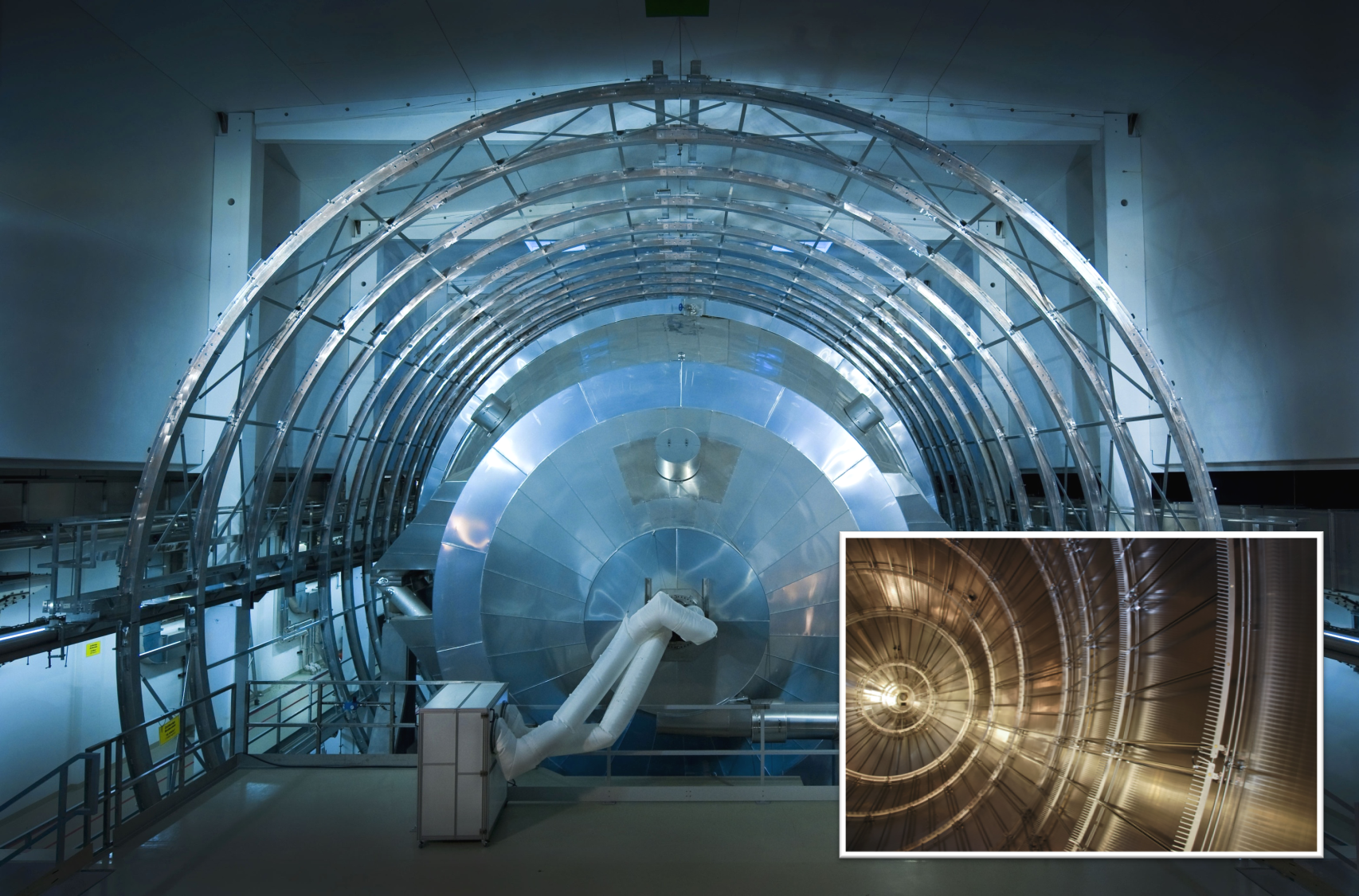}}
  \hspace{0.2cm}
  \subfigure[]{\includegraphics[width = 0.38\textwidth]{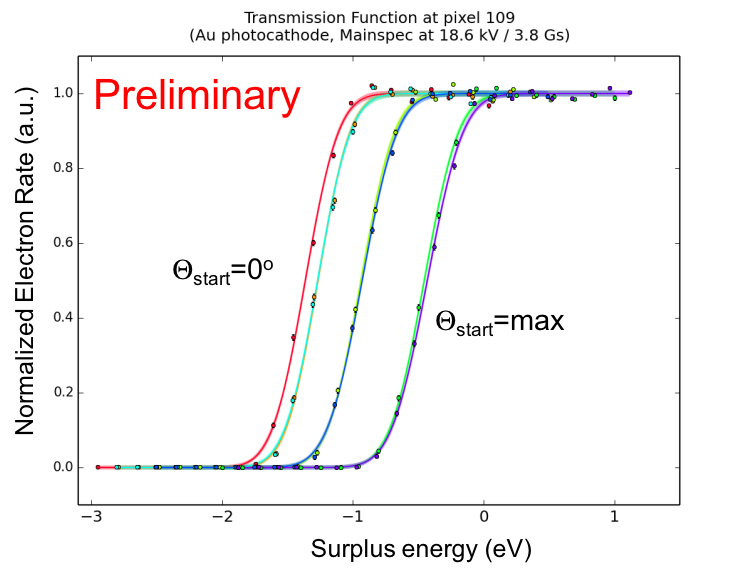}}
  \caption{a: Photograph of the KATRIN main spectrometer surrounded by its large air coil system used to fine-shape the magnetic field. The inset depicts the inner surface of the spectrometer, which is equipped with the inner electrode system to fine-tune the retarding potential. b: Preliminary transmission function. The larger the starting angle of the electrons, the more surplus energy they need to overcome the retarding potential. The shift of the transmission function for $0^{\circ}$ and maximal angle determines the energy resolution of the spectrometer.}
 \label{fig:trans}
\end{figure}

During two commissioning phases in 2013--2015 the background and transmission properties of the main spectrometer and focal plane detector~\cite{fpd} were studied. The transmission measurements, performed with an angular-selective electron gun, revealed an excellent energy resolution and confirmed that the spectrometer is working as a MAC-E-filter as expected, see figure~\ref{fig:trans}~\cite{Groh}. The anticipated radon-induced background~\cite{Mer13} could be reduced to a negligible level making use of a liquid-nitrogen-cooled baffle system. However, a remaining background level of $\sim$~100~mcps (as opposed to desired 10~mcps), is still under investigation~\cite{Harms}.

The neutrino mass measurement will prospectively start in 2017. With the start of the measurement, the sensitivity of KATRIN improves rapidly reaching the sub-eV level already after a few months of measurement time. After three years of data taking (5 calendar years) a balance between statistical and systematic error is reached. At this point, a 5$\sigma$ discovery level of $m_{\nu_e}=350$~meV and a 90\% upper limit of $m_{\nu_e}=200$~meV is attained. 

\subsection{The Project 8 Experiment}
Project 8 is exploring a new technique for $\beta$-spectrometry based on cyclotron radiation~\cite{p8_idea}. Using molecular tritium this approach could in principle reach the same sensitivity as the KATRIN experiment, but with quite different systematic uncertainties. 

\subsubsection{Working principle}
The general idea of this technique is to measure the coherent electromagnetic cyclotron radiation of the $\beta$-electron. As opposed to KATRIN, where the electron has to be extracted from the gaseous tritium source to measure its energy, here, the tritium source is transparent to the cyclotron radiation. The cyclotron frequency depends on the kinetic energy via the relativistic $\gamma$ factor. 

The technical realization of this approach consists of a magnetic trap inside of a wave guide. The magnetic field determines the frequency range of the $\beta$-electrons and the wave guide dimensions are chosen accordingly to match the frequency band of interest. For 18.6-keV electrons in a 1-T magnetic field the cyclotron frequency is 27.009~GHz. 

The radiated power scales with $B^2$ and $\sin^2\theta$, where $\theta$ is the angle between the momentum vector of the electron and the direction of the magnetic field. Hence, large angles and a sufficiently strong magnetic field are required. For an electron with an energy near the tritium endpoint, approximately 1.2~fW is radiated in a 1-T magnetic field at a pitch angle of $90^{\circ}$. By choosing a magnetic field setting with a very shallow trap, the pitch angle spread and magnetic field inhomogeneity can be reduced, which improves the energy resolution.

\subsubsection{Status and Sensitivity}
With a first prototype setup, the Project 8 collaboration successfully provided a proof-of-principle of the new technique~\cite{p8_signal}. The wave guide of 10.7~$\times$~4.3~$\mathrm{mm}^2$ cross section and 7.6~cm length was placed inside a warm bore magnet of about 1~T. An additional coil operated with a current of up to 2~A provided a shallow magnetic trap of -8.2~mT depth, that confined all electrons with pitch angles larger than $85^{\circ}$. 

\begin{figure}
  \centering
  \subfigure[]{\includegraphics[width = 0.41\textwidth]{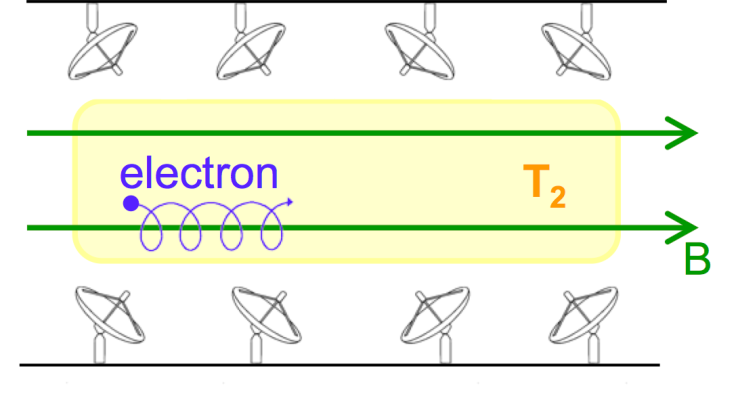}}
  \subfigure[]{\includegraphics[width = 0.41\textwidth]{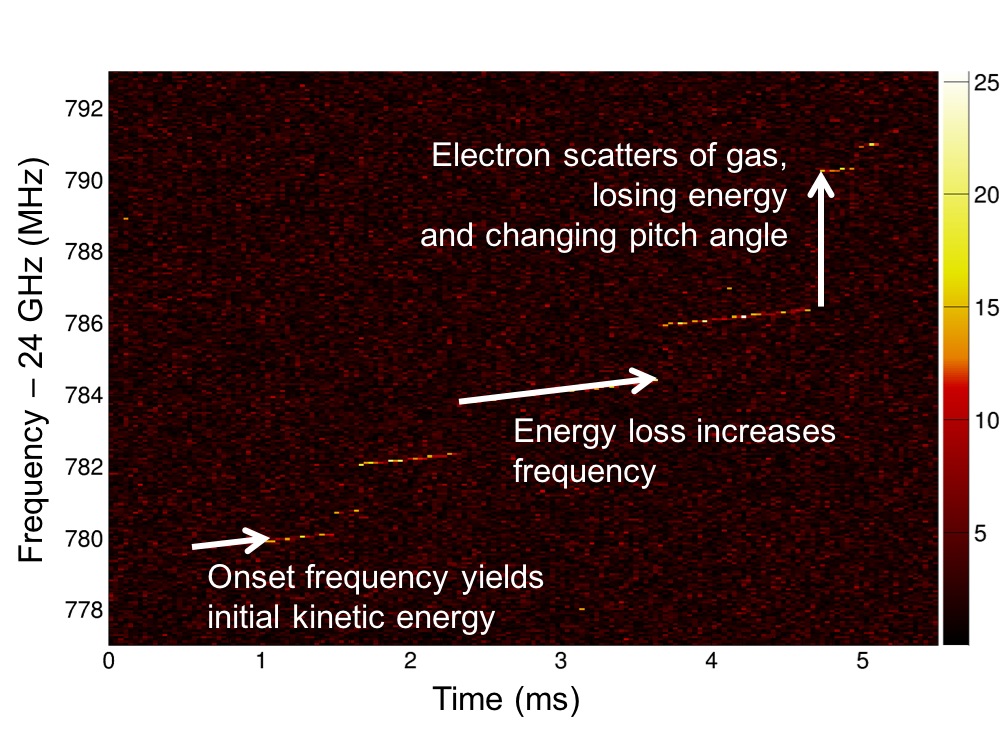}}
  \caption{a: Basic working principle of the Project 8 experiment. Electrons from tritium $\beta$-decay are trapped in magnetic field. Their relativistic cyclotron radiation is detected by a wave guide (here depicted as antenna array) b: First detection of a single electron via cyclotron radiation. The onset of the frequency yields the initial energy of the $\beta$-electron. As it  radiates and scatters it looses energy and hence increases its cyclotron frequency. }
 \label{fig:P8}
\end{figure}

For test measurements the cell was filled with krypton gas. ${}^{83m}$~Kr is a meta-stable state which decays via internal conversion processes emitting electrons in the keV-energy range. Figure~\ref{fig:P8}b shows the signature of a trapped electron. By reducing the depth of the trap an impressive energy resolution of FWHM(@30.4~keV) =  15~eV could be reached. The collaboration is currently aiming to use the prototype setup in conjunction with tritium to test its performance for a continuous energy spectrum. At the same time the options for scaling-up the setup and the usage of atomic tritium are investigated.

Preliminary and optimistic sensitivity studies~\cite{p8_sensi} show that with $\sim$1 year of data taking, with a density of  $10^{11}$ molecules/$\mathrm{cm}^{3}$ and a sensitive volume of 10 cubic meter a sensitivity of $ m_{\nu_e}\approx100$~meV (90\%C.L.) could be reached. This is the intrinsic limit dictated by the energy broadening due to the molecular final state distribution. An instrument with an atomic tritium source of $10^{12}$  atoms/$\mathrm{cm}^{3}$ and a sensitive volume of 100 cubic meters could, in principle reach a sensitivity of $m_{\nu_e}=40$~meV.

\subsection{Electron Capture on Holmium}
Currently, three experiments explore the approach of using electron capture on ${}^{163}$Ho to probe the neutrino mass: ECHo, HOLMES, and NuMECS. These experiments are complementary to tritium-based techniques both from a technical point-of-view and the fact that in this case the effective electron neutrino (as opposed to anti-neutrino) mass is measured.

\subsubsection{Working principle}
The basic idea is to place the ${}^{163}$Ho source inside an absorber material with low heat capacity. X-rays and electrons emitted in the de-excitation of the ${}^{163}$Dy$^{*}$ daughter atom create phonons in the absorber material and cause a small temperature increase. This temperature change is detected by ultra-sensitive thermometers such as transition edge sensors (TES) or magnetic metallic calorimeters (MMC). 

The calorimetric concept avoids a number of systematic effects as compared to the MAC-E-filter technology. In particular energy losses due to scattering during the extraction of the electron from the gaseous tritium source are completely circumvented. Furthermore, the intrinsic energy broadening due to the final state distribution of molecular tritium is not present. However, the micro-calorimetric technique involves a different class of systematic effects and technical challenges. 

As opposed to the KATRIN experiment where only the electrons of the ROI are considered, in these experiments every single decay is detected. The total decay rate is typically twelve orders of magnitudes higher than the decay rate only in the last few eVs away from the endpoint. Hence, pile-up becomes a serious concern. To limit pile-up 1) a fast rise time is needed and 2) the source needs to be spread over a large number of detectors. To operate a large number of detectors in a cryogenic environment, however, a sophisticated multiplexed read-out technology is necessary. 

Compared with the well-understood super-allowed tritium $\beta$-decay, the theoretical description of the ${}^{163}$Ho  spectrum is still a challenge. This topic is addressed by several groups, who found that two- and three-hole excitations due to shake-up and -off processes need to be included and might significantly change the expected statistics at the endpoint~\cite{fas15, ruj15}.

\begin{figure}
  \centering
  \subfigure[]{\includegraphics[width = 0.41\textwidth]{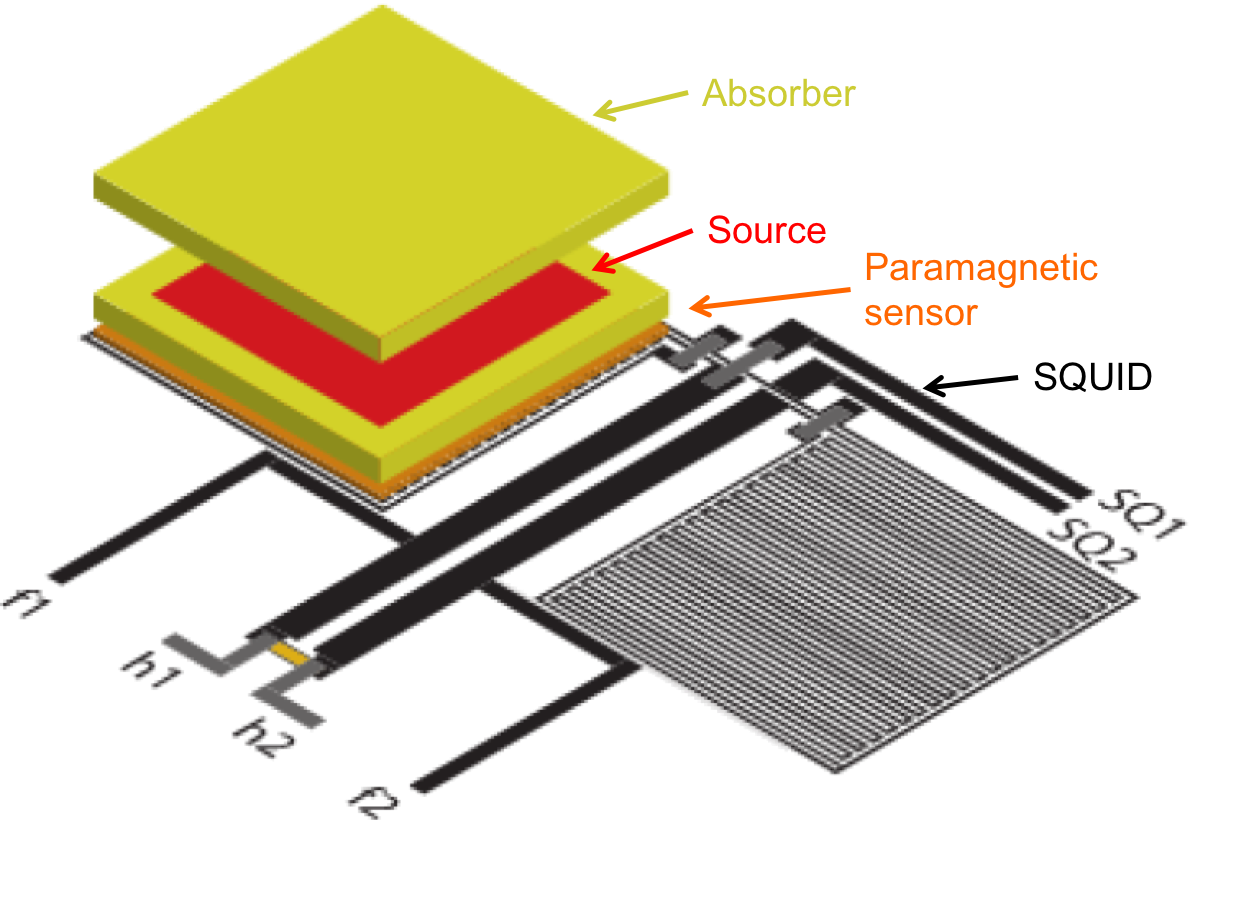}}
  \hspace{0.2cm}
  \subfigure[]{\includegraphics[width = 0.38\textwidth]{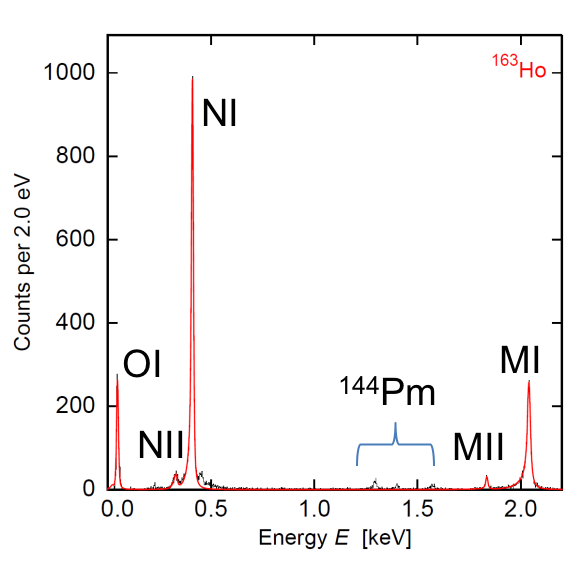}}
  \caption{a: Experimental setup of a micro-calorimetric detector. The source (red) is enclosed by a gold absorber (yellow). The paramagnetic temperature sensor (orange) is read-out by a SQUID system~\cite{echo_spectrum}. b:  ${}^{163}$Ho spectrum measured by the ECHo collaboration. This spectrum presents is the first calorimetric measurement of the OI-line~\cite{echo_spectrum_future}}
 \label{fig:Kink}
\end{figure}

\subsubsection{Status and Sensitivity}
Three groups are currently developing neutrino mass experiments based on electron-capture on ${}^{163}$Ho:

ECHo~\cite{echo} is using MMCs for read-out. The holmium source is enclosed in a gold absorber, which is attached to an Au:Er paramagnetic sensor at 30~mK. The temperature change causes a drop of the magnetization of the sensor which is detected by a SQUID. With their first prototype ECHo could demonstrate excellent energy resolution of 7.6~eV @ 6~keV and fast rise times of $\tau = 130$~ns. A larger detector array with 16 pixels, increased purity, and activity (0.1~Bq) is being tested at the moment. 

HOLMES~\cite{holmes} is making use of the TES technology. The collaboration is currently performing detector and read-out R\&D. In particular, a custom ion-implanter is being assembled in Genova to embed the ${}^{163}$Ho in the detectors. The first test with a ${}^{163}$Ho source will prospectively begin in 2017.

NuMECS~\cite{numecs} is also pursuing the TES technology. This group's focus is on ${}^{163}$Ho production via  proton activation of dysprosium, as opposed to the more common neutron irradiation on ${}^{162}$Er. With their prototype where the source was enclosed as liquid drop in a nanoporous gold absorber, NuMECS successfully measured a ${}^{163}$Ho spectrum with an energy resolution of about 40~eV FWHM.

A total statistics of $10^{14}$ events is needed to reach a sub-eV sensitivity. Assuming a rise time that allows for 10~Bq per detector, $10^5$ detectors are needed to reach $m_{\nu_e}=1$~eV sensitivity within one year of measurement time.

\section{Conclusion}
The kinematics of $\beta$-decay provides a unique, model-independent means to measure the absolute neutrino mass. KATRIN will start taking data in the near future reaching a final sensitivity of 200~meV (90\%C.L.) after 3 years of data collection. The Project 8 collaboration proved a completely novel concept of measuring the $\beta$-electron's energy via its cyclotron frequency and holmium-based cryogenic experiments are advancing to reach the sub-eV sensitivity. These new approaches will provide complementary results and may show a path towards exploring the hierarchical neutrino mass regime.


\end{document}

%% file: econfmacros.tex
\def\beq{\begin{equation}}
\def\eeq#1{\label{#1}\end{equation}}
\def\eeqn{\end{equation}}

\def\beqa{\begin{eqnarray}}
\def\eeqa#1{\label{#1}\end{eqnarray}}
\def\eeqan{\end{eqnarray}}

\let\bar=\overbar

\def\Dslash{\not{\hbox{\kern-4pt $D$}}}
\def\dslash{\not{\hbox{\kern-2pt $\del$}}}

\def\msb{{\bar{\ssstyle M \kern -1pt S}}}

